\documentclass[11pt]{article}
\usepackage{amsmath}
\usepackage{array}
\usepackage{color}
\usepackage{dcolumn}
\usepackage{graphicx}
\usepackage[figuresright]{rotating}

\def \ppnp#1#2#3{{\it Prog. Part. Nucl. Phys.} {\bf#1} (#3) #2}
\def \prd#1#2#3{{Phys. Rev. D} {\bf#1}, #2 (#3)}
\def \prl#1#2#3{{Phys. Rev. Lett.} {\bf#1}, #2 (#3)}

\begin{document}

\begin{center}
{\Large\bf Bottomonium(-like) State Spectroscopy at B-Factories}
\end{center}

\vspace{3mm}

\begin{center}
G.~Tatishvili \footnote{On behalf of the Belle Collaboration} \\
Pacific Northwest National Laboratory\\
902 Battelle Boulevard, Richland, WA, USA
\end{center}

\vspace{3mm}

\begin{abstract}
Bottomonium spectroscopy is a key source necessary for understanding of Quantum Chromodynamics. The expected results of this endeavor will provide important tests for various theoretical approaches to understanding quark-antiquark interaction dynamics.
Recent results in bottomonium spectroscopy are presented. 
\end{abstract}

\vspace{3mm}

The first evidence for bottomonium, the bound states of $b$ and anti-$b$ quarks, was observed in the spectrum of $\mu^+\mu^-$ pairs produced in 400 GeV 
proton-nucleus collisions at Fermilab~\cite{herb,innes}. In the $e^+e^-$ collision process the entire collision energy of the initial $e^+e^-$ turns into the rest mass of the 
$\Upsilon$ state and the beam energy must be matched to the resonance mass.  Figure 1 shows the bottomonium bound states including newly discovered states the spin-singlet states $\eta_b(1S, 2S)$ and $h_b(1P, 2P)$, the first D-wave states,  one or more candidates for spin-triplet $\chi_{bJ} (3P)$ excitations, and above-threshold states with transitions to states below threshold~\cite{todd,gocha}.

Major contributions in the study of bottomonium were made by experiments at the electron-positron colliders KEKB (Belle Detector), 
PEP-II (BaBar Detector).

Belle and BaBar detectors were large-solid-angle magnetic spectrometers that consisted of Drift Chambers, silicon vertex detectors, 
electromagnetic calorimeter and superconducting solenoid that provided a 1.5 T magnetic field.
BaBar collected largest dataset of $\Upsilon(3S)$. Experiment Belle collected largest  $\Upsilon(1S)$,  $\Upsilon(2S)$,  $\Upsilon(4S)$,  $\Upsilon(5S)$ samples. 

The spin-singlet states $h_b(nP)$ provide information about the hyperfine interaction in bottomonium. Measurements of the $h_b(nP)$ masses gives access to the P-wave hyperfine splitting between the spin-weighted average mass of the P-wave triplet states $\chi_{bJ}(nP)$ and that of the corresponding $h_b(nP)$.
The hyperfine splitting for bottomonium states is proportional to the square of the wave function at the origin, which is expected to be
nonzero for zero orbital angular momentum (L) of quark anti-quark system. For L=1, the mass difference between the spin singlet and spin-average mass of the triplet state is expected to be zero.

\begin{figure}
  \begin{center}
    \begin{minipage}[t]{1.0\linewidth}
	\includegraphics[width=90mm, height=60mm]{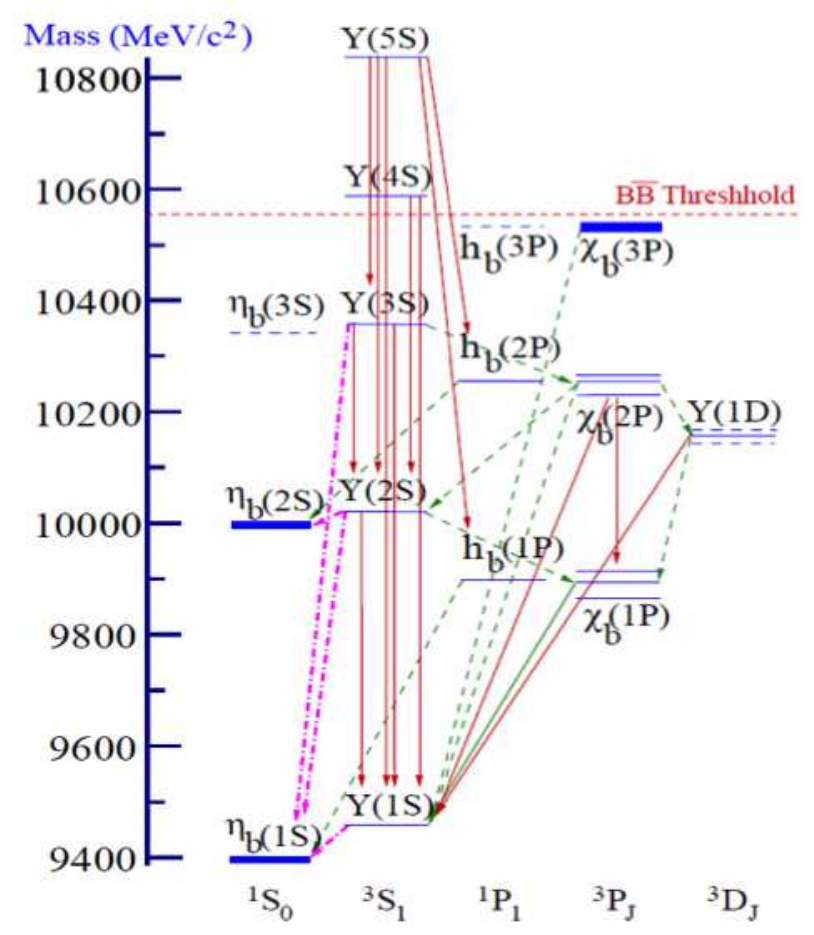} 
    \end{minipage}\hfill
    \caption{ Current knowledge of the bottomonium system. Solid lines correspond
to known states while dashed lines are predicted ones. The thicker lines indicate
the range of measured masses for newly discovered states. (Solid, dashed, dotdashed)
arrows denote (hadronic, electric dipole [E1], and magnetic dipole [M1])
transitions, respectively~\cite{todd}}
  \end{center}
\end{figure}

Experiment Babar reported observation of $\approx$3 standard deviation excess of events above background in the recoil mass distribution against the $\pi^0$ at mass  9902$\pm$4(stat.)$\pm$2(syst.) MeV/c$^2$ in the transition $\Upsilon{(3S)}\rightarrow \pi^0 h_b{(1P)}\rightarrow \pi^0\gamma\eta_b(1S)$~\cite{hbbha}. 
This excess was consistent with the expectation for the $h_b(1P)$ bottomonium state.

The first significant signal for this state, however, came from the transitions $\Upsilon(5S)\rightarrow h_b(nP)\pi^+\pi^-$, in the experiment Belle~\cite{hbbelle}. The $h_b{(nP)}$ states were produced via $e^+e^-\rightarrow h_b{(nP)}\pi^+\pi^-$  and observed in the $\pi^+\pi^-$ missing mass spectrum of hadronic events. Figure 2 shows the background subtracted inclusive $M_{miss}$ spectrum. 

The measured masses of $h_b{(1P)}$ and $h_b{(2P)}$ states were $M=(9899.1\pm 0.4 \pm 1.0)$ MeV/c$^2$ and $M=(10259.8\pm 0.5\pm 1.1)$ MeV/c$^2$, respectively.  Using the measured $h_b$ and world average masses of $\chi_{bJ}(nP)$ states Belle determined the hyperfine splittings to be $\Delta{M}_{HF} = (+0.8 \pm 1.1)$ MeV/c$^2$  and $(+0.5\pm 1.2)$ MeV/c$^2$ for $h_{b}(1P)$ and $h_{b}(2P)$, respectively~\cite{etab}.

\begin{figure}
  \begin{center}
    \begin{minipage}[t]{1\linewidth}
	\includegraphics[width=125mm, height=40mm]{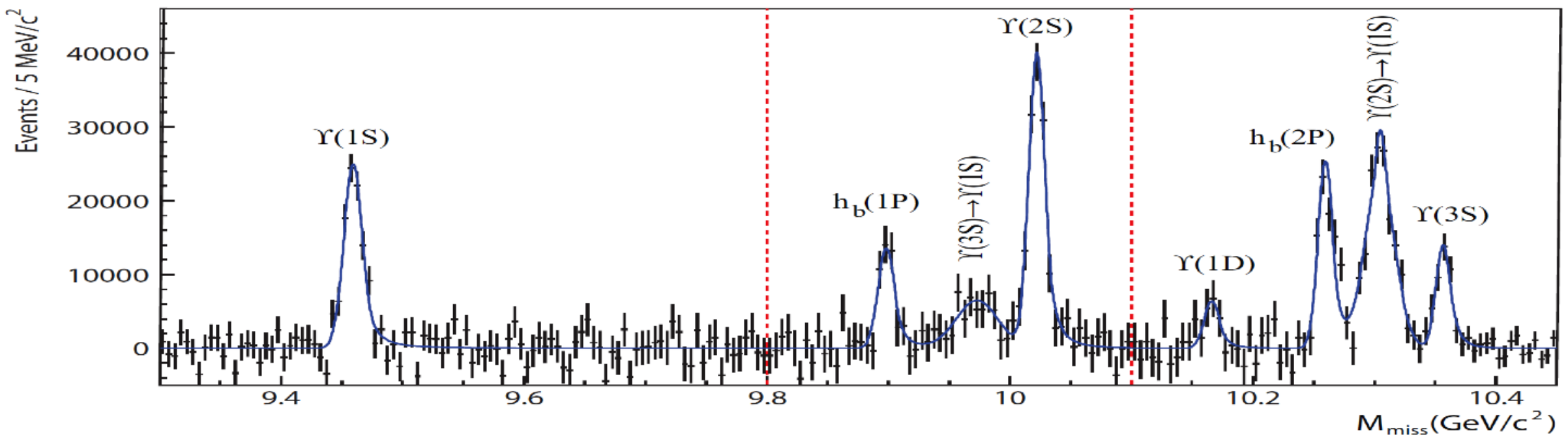}  
    \end{minipage}\hfill
    \caption{The background subtracted inclusive $M_{miss}$ spectrum (points with errors).  The vertical lines indicate boundaries of the fit regions. Overlaid smooth curve is the resulting fit function.}
  \end{center}
\end{figure}
\newpage

Experiment Belle determined the ratio of cross sections $\sigma (h_b(nP)\pi^+\pi^-)/$  $\sigma (\Upsilon(2S)\pi^+\pi^-)$ to be  0.45 $\pm$ 0.08${^{+0.07}_{-0.12}}$ for the $h_b(1P)$ and 0.77$\pm$0.08${^{+0.22}_{-0.17}}$ for the $h_b(2P)$. Hence,  
$\Upsilon{(5S)}\rightarrow h_b{(nP)}\pi^+\pi^-$ and $\Upsilon{(5S)}\rightarrow \Upsilon{(2S)}\pi^+\pi^-$ 
proceed at similar rates despite the fact that the production of  $h_b(nP)$ requires a spin flip of a $b$ quark~\cite{hbbelle}.

The Belle Collaboration reported results of study of resonant substructure in the decays of  $\Upsilon{(5S)}\rightarrow \Upsilon{(nS)}\pi^+\pi^-$ $(n = 1,2,3)$ and $\Upsilon(5S)\rightarrow h_b(mP)\pi^+\pi^-$ $(m = 1,2)$~\cite{hbsubc}.
Figure 3 show the $h_b(1P)$ (left plot) and $h_b(2P)$ (right plot) yields as a function of the missing mass recoiling against the pion. 

\begin{figure}
  \begin{center}
    \begin{minipage}[t]{1\linewidth}
	\includegraphics[width=125mm, height=50mm]{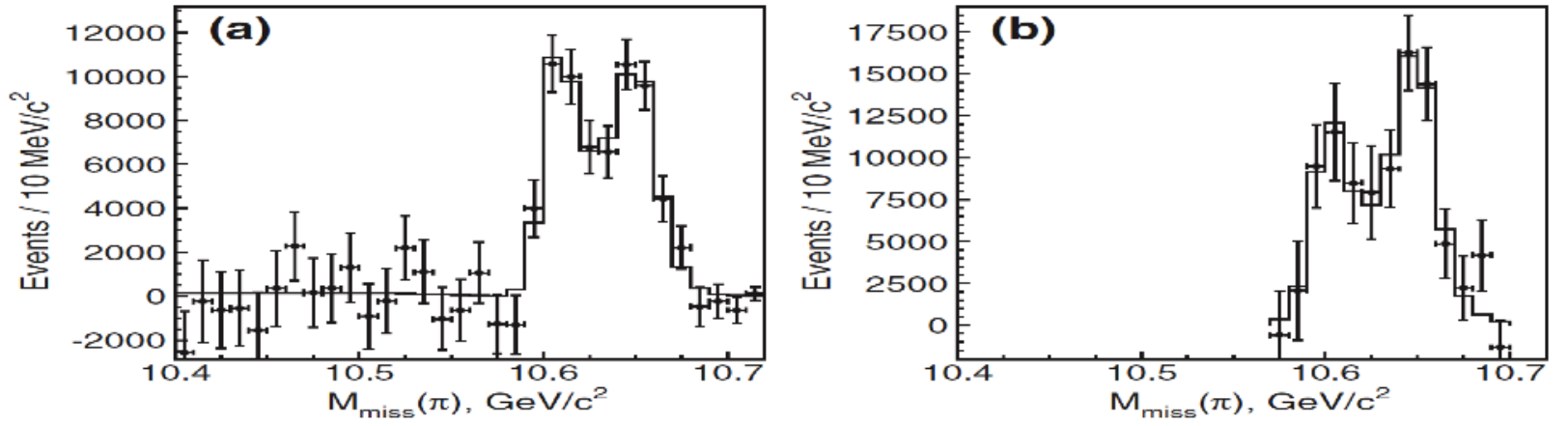}  
    \end{minipage}\hfill
    \caption{The $h_b(1P)$  (a) and $h_b(2P)$  (b) yields as a function of the missing mass recoiling against the pion. Fit results are presented as a histograms.}
  \end{center}
\end{figure}

A clear two-peak structure indicated the production of $Z_b(10610)$ and $Z_b(10650)$ states. In total Collaboration Belle observed two $Z_b(10610)$ and $Z_b(10650)$ bottomonium-like  resonances in five different decay channels $\Upsilon{(nS)}\pi^\pm$  $(n=1,2,3)$ and $h_b{(mP)}\pi^\pm$  $(m=1,2)$.   The Collaboration Belle also reported the first observation of $\Upsilon{(5S)}\rightarrow \Upsilon (2S) \pi^0\pi^0$ decays~\cite{hbsubn}. Evidence for the $Z^0_b{(10610)}$ with 4.9$\sigma$ significance was found in a Dalitz plot analysis of $\Upsilon{(5S)}\rightarrow \Upsilon{(2S)}\pi^0\pi^0$ decays. The results are obtained with a 121.4 fb$^{-1}$ data sample collected with the Belle detector at the $\Upsilon{(5S)}$ resonance. Figure 4 show comparison of the fit results (open histograms) with experimental data (points with error bars). Left plot shows $\Upsilon{(2S)} \pi^0 \pi^0$ events in the signal region; Right plot - $\Upsilon{(1S)} \pi^0 \pi^0$ events.

The minimal quark content  of the $Z_b(10610)$ and $Z_b(10650)$ states requires a four quark combination.  The measured masses of these states are a few MeV/c$^2$ above the thresholds for the open beauty channels  which suggests a "molecular" nature of these states i.e. their internal dinamics is dominated by the coupling to meson pairs  $B^*\bar{B} - B\bar{B}^*$  and $B^*\bar{B}^*$~\cite{bbar}.

\begin{figure}[!htb]
\begin{minipage}[t]{62mm}
	\includegraphics[width=62mm, height=60mm]{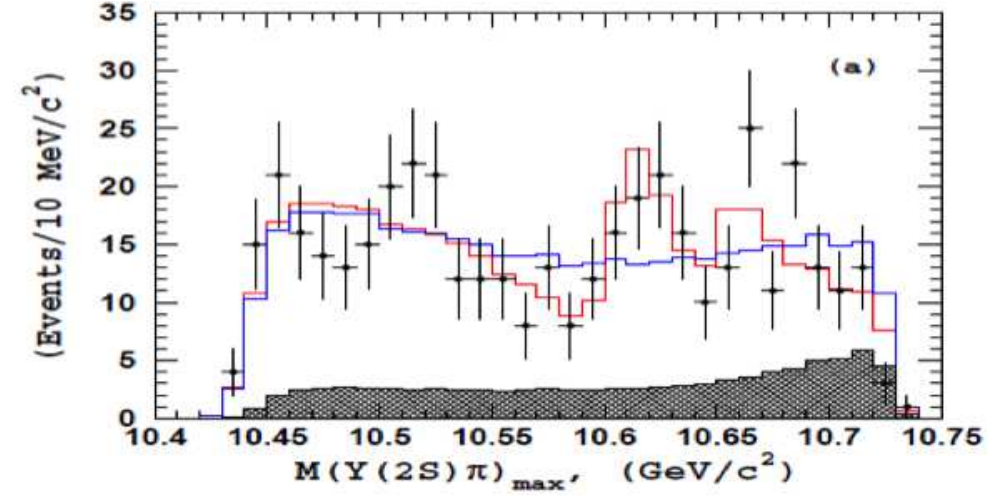} 
\end{minipage}
\hfill
\begin{minipage}[t]{62mm}
	\includegraphics[width=62mm, height=60mm]{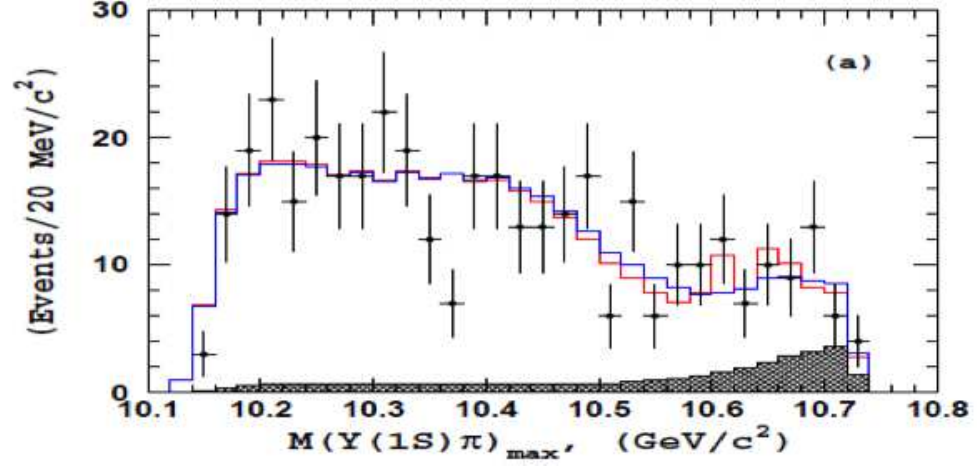} 
\end{minipage}
\caption{Comparison of the fit results (open histograms) with experimental data (points with error bars). Left plot shows $\Upsilon{(2S)} \pi^0 \pi^0$ events in the signal region; Right plot - $\Upsilon{(1S)} \pi^0 \pi^0$ events.}
\end{figure}

The mass of bottomonium state $\eta_b(1S)$, was predicted 35 - 100 MeV below the $\Upsilon{(1S)}$ mass by a variety of potential models.
The $\Upsilon{(1S)}\rightarrow \gamma \eta_b(1S)$ transition leads to a very low energy photon which is not too easy  to distinguish from the background photons produced due to transition $\Upsilon{(1S)}\rightarrow \pi^0 X\rightarrow \gamma\gamma X$. More energetic photon can be observed in the transition
of $\Upsilon{(3P)}\rightarrow \gamma \eta_b(1S)$. 

\begin{figure}
  \begin{center}
    \begin{minipage}[t]{1\linewidth}
	\includegraphics[width=90mm, height=50mm]{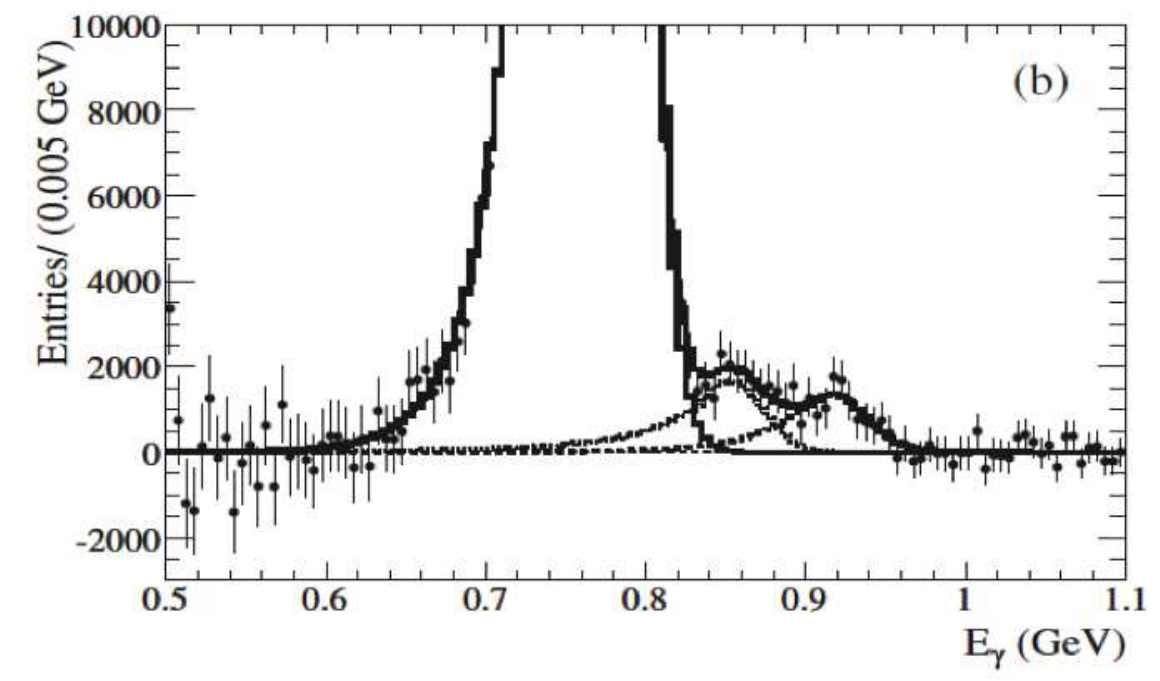}  
    \end{minipage}\hfill
    \caption{Inclusive photon spectrum after subtracting the nonpeaking background, with PDFs for the $\chi_{bJ}(2P)$ peak (solid line), ISR $\Upsilon{(1S)}$ dotted line), $\eta_b$ signal (dashed line) and the sum of all three (solid line).}
  \end{center}
\end{figure}

Collaboration BaBar reported the results of a search for the bottomonium ground state $\eta_b{(1S)}$~\cite{etabba} in the photon energy spectrum 
with a sample of 109 million $\Upsilon{(3S)}$ recorded at the PEP-II B factory at SLAC.  Observed  peak in the photon energy spectrum at 
E=921.2${^{+2.1}_{-2.8}}$(stat.)$\pm$2.4(syst.) MeV was interpreted as being due to monochromatic photons from the radiative transition $\Upsilon{(3S)}\rightarrow \gamma \eta_b{(1S)}$.  
Figure 5 shows Inclusive photon spectrum after subtracting the nonpeaking background. $\eta_b$ signals are presented by dashed line.
Observed $\eta_b$ mass was 9388.9${^{+3.1}_{-2.3}}$(stat.)$\pm$2.7(syst.) MeV/c$^2$. The hyperfine mass splitting ($\Upsilon{(1S)}$ - $\eta_b{(1S)})$ was 71.4${^{+2.3}_{-3.1}}$(stat.)$\pm$2.7(syst.) MeV/c$^2$.

CLEO weakly confirmed~\cite{etabcle} with 6x10$^6$ $\Upsilon{(3S)}$ the BaBar result.

Belle's large $h_b$ sample provide an opportunity to study the $\eta_b(1S)$ and $\eta_b(2S)$ ground state of the bottomonium system. 
Belle reported the first observation of the radiative transition $h_b{(1P)}\rightarrow \eta_b{(1S)}\gamma$, where the $h_b{(1P)}$ was produced in 
$\Upsilon{(5S)}\rightarrow h_b{(1P)}\pi^+\pi^-$ dipion transitions~\cite{etab}. Figure 6 (a) and (b) show the radiative transitions from the $h_b{(1P)}$ and $h_b{(2P)}$ to 
the $\eta_b{(1S)}$. First evidence for the $\eta_b{(2S)}$ observed using radiative transition from the $h_b{(2P)}$ is presented on the Fig. 6 (c).
Experiment Belle measured the $\eta_b{(1S)}$ mass to be
(9401.0$\pm$1.9${^{+1.4}_{-2.4}}$)MeV/c$^2$ with a decay branching fraction of $B[h_b{(1P)}\rightarrow \eta_b{(1S)}\gamma]$ = (49.8$\pm$6.8$^{+10.9}_{-5.2}$)\%. 
The measured $\eta_b{(1S)}$ mass corresponds to a hyperfine splitting of (59.3$\pm$1.9$^{+2.4}_{-1.4}$) MeV/c$^2$. This value deviates significantly from the current world average obtained from measurements of $\Upsilon{(3S)}\rightarrow \eta_b{(1S)}\gamma$ and $\Upsilon{(2S)}\rightarrow \eta_b{(1S)}\gamma$ reactions.

QCD multipole expansion formalism described hadronic transitions between bottomonia. 
The $\eta$ and $\pi^0$ transitions between vector bottomonia should be mediated either by two M1 gluons or by one E1 and one M2 gluon; both cases
imply a spin flip of the $b$ quark. The corresponding total amplitude should scale as $1/m_b$, and its measurement yields
information about the chromomagnetic moment of the $b$ quark~\cite{etatran}.

The transition $\Upsilon{(2S)}\rightarrow \eta \Upsilon{(1S)}$ was first observed by the CLEO Collaboration~\cite{etacle} in about 9 million $\Upsilon{(2S)}$. The BaBar Collaboration essentially confirmed this result~\cite{etabab}. 

Collaboration Belle reported results of rare hadronic transitions $\Upsilon{(2S)}\rightarrow \Upsilon{(1S)}\eta$ and  
$\Upsilon{(2S)}\rightarrow \Upsilon{(1S)} \pi^0$ using a sample of 158 x 10$^6$ $\Upsilon{(2S)}$ decays~\cite{etatran}.
Figure 7 (left plot) shows sum of mass distribution of the $\eta$ candidates in $\gamma\gamma\mu^+\mu^-$, $\gamma\gamma e^+ e^-$, $\pi^+\pi^0\pi^-\mu^+\mu^-$ and $\pi^+\pi^0\pi^- e^+ e^-$ final states. The distribution of the $\gamma\gamma$ pair invariant mass for 
$\Upsilon{(2S)}\rightarrow \Upsilon{(1S)}\pi^0$ candidates are presented on Fig. 7 (right plot).
Experiment Belle obtained the value of ratio: $R_{\eta},_{\pi^+\pi^-}$ = $B(\Upsilon{(2S)}\rightarrow \Upsilon{(1S)}\eta)/B(\Upsilon{(2S)}\rightarrow \Upsilon{(1S)}\pi^+\pi^-)$ = (1.99$\pm$0.14(stat.)$\pm$0.11(syst.)x10$^{-3}$ and $B(\Upsilon{(2S)}\rightarrow \Upsilon{(1S)}\pi^0)/B(\Upsilon{(2S)}\rightarrow \Upsilon{(1S)}\pi^+\pi^-)$ $<$ 2.3 x 10$^{-4}$. The value of $R_{\eta},_{\pi^+\pi^-}$ is 14\%  below the value extracted from~\cite{volo} .
Assuming the branching fraction $B(\Upsilon{(2S)}\rightarrow \Upsilon{(1S)}\pi^+\pi^-)$ = (17.92$\pm$0.26)\%~\cite{beri}, a new measurement of 
$B(\Upsilon{(2S)}\rightarrow \Upsilon{(1S)}\eta)$ was obtained: $B(\Upsilon{(2S)}\rightarrow \Upsilon{(1S)}\eta)$ = (3.57$\pm$0.25(stat.)$\pm$0.21(syst.)x10$^{-4}$.

\begin{figure}
  \begin{center}
    \begin{minipage}[t]{1\linewidth}
	\includegraphics[width=125mm, height=40mm]{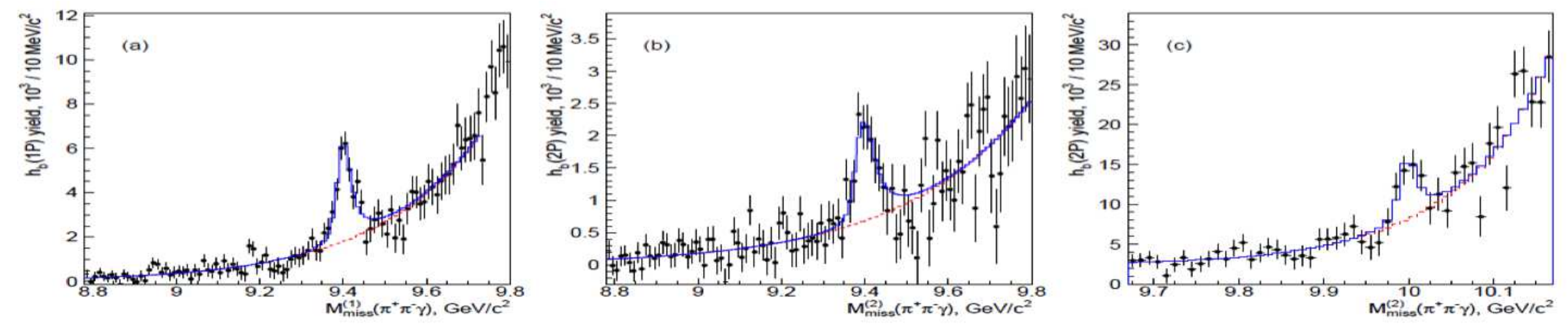}  
    \end{minipage}\hfill
    \caption{The $h_b(1P)$ yield (left) and the $h_b(2P)$ yield in the $\eta_b(1S)$ region (center) and in the $\eta_b(2S)$ region (right). The solid (dashed) histogram is the fit result (background component of the fit function). }
  \end{center}
\end{figure}
\newpage

\begin{figure}[!htb]
\begin{minipage}[t]{62mm}
	\includegraphics[width=62mm, height=50mm]{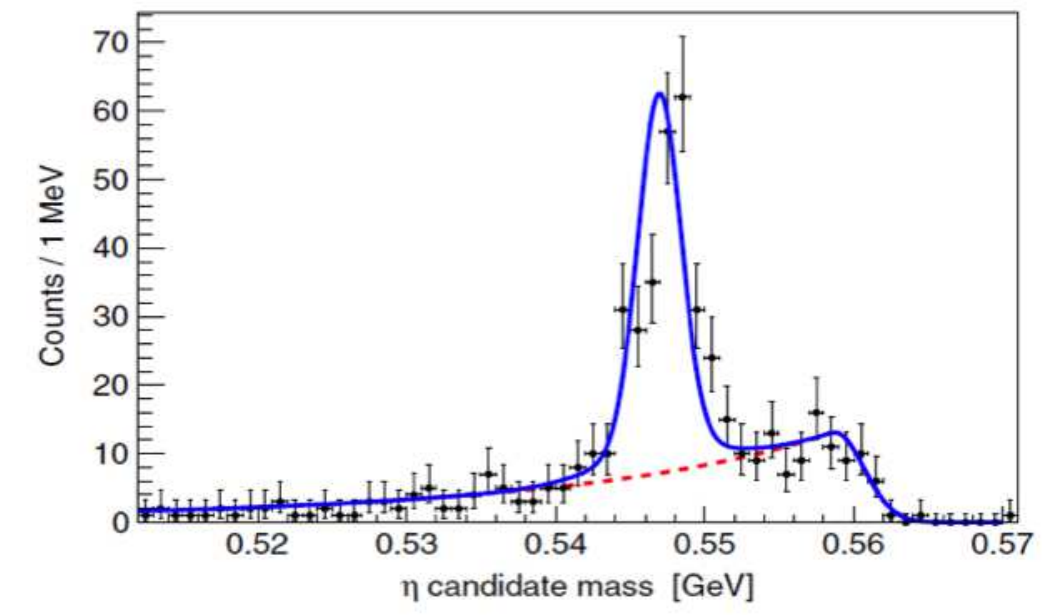} 
\end{minipage}
\hfill
\begin{minipage}[t]{62mm}
	\includegraphics[width=62mm, height=50mm]{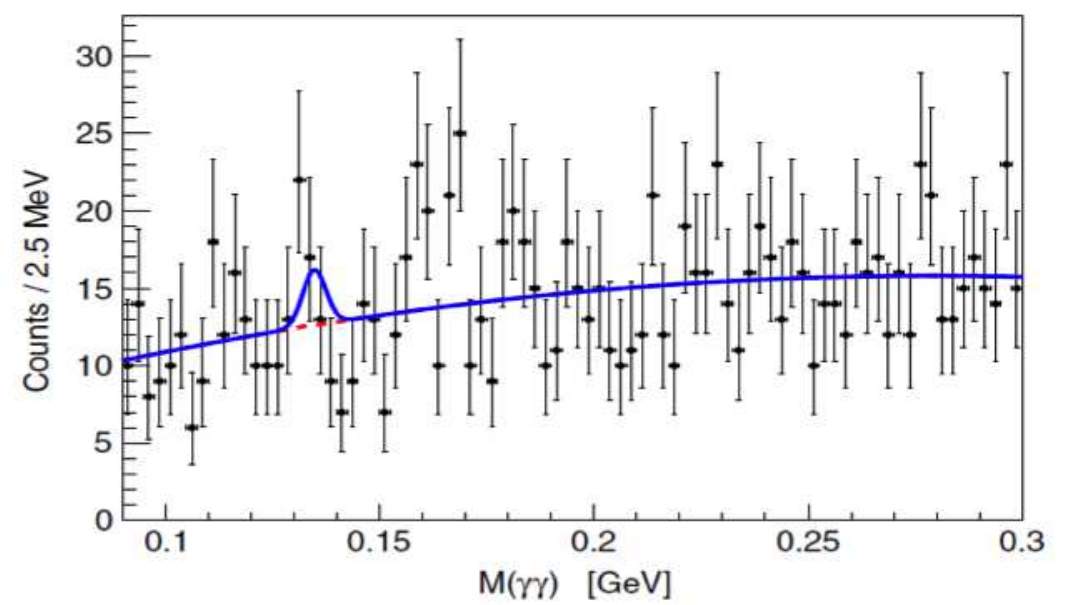} 
\end{minipage}
\caption{$\gamma\gamma$/$\pi^+\pi^-\pi^0$ invariant mass for $\Upsilon(2S) \rightarrow \Upsilon(1S) \eta$ candidates, summing 
all the four final states. The fit function in blue, solid represents the best fit with the red dashed curve showing its the background component (left plot). Fit to the $\gamma\gamma$ pair invariant mass for $\Upsilon(2S)\rightarrow \Upsilon(1S)\pi^0$ candidates (right plot). The fit function in blue, solid represents the best fit; the red dashed curve shows its background component.}
\end{figure}

\end{document}